\documentclass[aps,pre,floatfix,superscriptaddress,twocolumn,amsfonts,amssymb,letterpaper]{revtex4}
\usepackage[utf8]{inputenc}
\usepackage{graphicx}
\usepackage{amsmath}
\usepackage{mathtools}

\newcommand{\av}[1]{\left\langle#1\right\rangle}

\newcommand{\dmu}{\Delta\mu}
\newcommand{\w}{w}

\newcommand{\T}{{\mathcal T}}

\newcommand{\Li}{\operatorname{Li}}
\newcommand{\OV}{\Omega_V}
\newcommand{\Omu}{\Omega(\mu)}
\newcommand{\pib}{\bar\pi}

\begin{document}

\title{Long-range disassortative correlations in generic random trees} 

\author{Piotr Bialas}
\email[]{pbialas@th.if.uj.edu.pl}
\affiliation{Marian Smoluchowski Institute of Physics, Jagellonian University, Reymonta 4, 30--059 Krakow, Poland}
\affiliation{Mark Kac Complex Systems Research Centre, Faculty of Physics, Astronomy and Applied Computer Science,\\ Jagellonian University, Reymonta 4, 30--059 Krakow, Poland}
\author{Andrzej K. Oleś}
\email[]{oles@th.if.uj.edu.pl}
\affiliation{Marian Smoluchowski Institute of Physics, Jagellonian University, Reymonta 4, 30--059 Krakow, Poland}

\begin{abstract}
  We explicitly calculate the distance dependent correlation functions
  in a maximal entropy ensemble of random trees. We show that
  correlations remain disassortative at all distances and vanish only
  as a second inverse power of the distance. We discuss in detail the
  example of scale-free trees where the diverging second moment of
  the degree distribution leads to some interesting phenomena.\\\\
\begin{tabular}{p{7.3cm} p{6.6cm}}
Published in: Phys. Rev. E {\bf 81}, 041136 (2010). &PACS numbers: 05.90.+m, 89.75.Hc, 05.10.--a
\end{tabular}
\end{abstract}

\maketitle

\section{Introduction}

The knowledge of correlations is important and interesting for any
system.  Looking from the practical point of view correlation means
additional information: if two quantities are correlated,
the knowledge of one of them implies certain information about the other one.  In
physical systems correlations usually indicate interactions
between parts of the system. The prototypical example is given by the
Ising spin system where the nearest-neighbors interactions induce
long-range correlations leading to a phase transition.

The situation in random graphs is somewhat different. It is known that
for random geometries even in the absence of any explicit terms
inducing interactions between vertices their degree may be
correlated \cite{bs,b1,b2,b3,Satorras2001,msz,pn,bo,kim}. Moreover, those
correlations are long-range, i.e., they fall off as some power of
distance \cite{bs,b1,b2,b3}. They are generated by model
constraints rather than by explicit interactions.  It should be also
stressed that the distance dependent correlation functions in the
ensemble of random graphs are much more complicated objects than their
fixed lattice counterparts \cite{bs,b2}. To see that let us take some
generic correlation function on random graphs
\begin{equation}
\av{\sum_{i,j}A({q_i}) B({q_j})\,\delta_{l,|i-j|}},
\end{equation}
where $q_i$ denotes the {\em degree} of the vertex $i$, i.e., the
number of branches emerging from it. $A$ and $B$ are some arbitrary
functions depending on the vertex degree and $|i-j|$ is the graph
(geodesic) distance between vertices $i$ and $j$. For random geometry
it makes no sense in general to choose two fixed points---that is why
we sum over all the pairs of points on the graph. The graph distance
is the length of the shortest path between those two vertices and as
such it is dependent on the whole graph. That means that the above
expression is not a two-point function but a highly nonlocal
object. That is a fundamental difference between random and fixed
geometries.

In this paper, we study in detail correlations between degrees
of vertices as a function of distance.  We consider an ensemble of
all labeled trees $\T(V)$ with a fixed number of vertices $V$, on
which we define the probability measure
\begin{equation}\label{eq:PT}
P(T)\equiv\Omega_V^{-1}\frac{1}{V!}\prod_{i\in T}\w_{q_i}.
\end{equation}
$\Omega_V$ denotes the partition function of this ensemble
(normalization factor) and $q_i$ is the degree of vertex $i$; $\w_{q}$'s $(q>0)$ are some
non-negative numbers (weights). 
This  is a maximal-entropy
ensemble with a given degree distribution 
(see Appendix). An important property of the measure
\eqref{eq:PT} is that it factorizes into a product of one-point
measures, so it does not introduce any explicit
correlations. This means that any observed correlations arise from the
fact that we consider a specific set of graphs and not from the
measure itself.  

We show that  the connected degree-degree correlations
are not zero and fall off with the distance as $l^{-2}$
\begin{eqnarray}\label{eq:pbcontree}
\pib^{con}_{q,r}(l)&=& \pib_{q,r}(l)-\pib_q(l)\pib_r(l)\nonumber\\
&=&-\frac{(q-2)(r-2)}{\left[2+ (\av{q^2}-4)(l-1)\right]^2}\,\pi_q\pi_r.
\end{eqnarray}
Here $\bar\pi_{q,r}(l)$ is the joint probability that two vertices
distance $l$ apart will have degrees $q$ and $r$, respectively. Those
correlations are disassortative.  The average degree of the distance $l$
neighbors of a vertex with degree $q$ decreases
\begin{equation}\label{eq:kl} 
\bar k_l(q)=
2+\frac{\av{q^2}-4}{q+(\av{q^2}-4)(l-1)}.
\end{equation}
For $l=1$ this reduces to the results obtained in Ref. \cite{kim}.  In the
following sections we provide the detailed definitions of the
quantities introduced above and derive those results. We will also
discuss what happens for the scale-free trees when $\av{q^2}$
diverges.
 
The paper is organized as follows: Sec. \ref{sec:correlations}
introduces some basic definitions concerning correlations in random
trees.  Then we derive the vertex degree distribution using the field
theory approach in Sec. \ref{sec:trees} and proceed on in
Sec. \ref{sec:correlationstrees} calculating the distance dependent
correlation functions. Two examples of Erd\"{o}s-R\'{e}nyi and
scale-free trees are given in Secs. \ref{sec:erdos} and \ref{sec:sf},
respectively.  In the following Sec. \ref{sec:simulations} the results
for scale-free trees are verified using Monte Carlo (MC) simulations.
Final discussion and summary of our results are given in
Sec.~\ref{sec:summary}.

\section{Correlations}
\label{sec:correlations}

For each graph we introduce
\begin{equation}
n_{q,r}(l)\equiv\sum_{i,j\in G}\delta_{q_i,q}\delta_{q_j,r}\delta_{|i-j|,l},
\end{equation}
which is the number of pairs of points with degrees $q$ and $r$ separated by the distance $l$. We
define two further quantities: the number of pairs at the distance $l$
with one end point of specified degree
\begin{equation}
n_q(l)\equiv\sum_{r}n_{q,r},
\end{equation}
and the number of all pairs of vertices at the distance~$l$
\begin{equation}
n(l)\equiv\sum_{q,r}n_{q,r}(l).
\end{equation}
If we want to define the joint probability $\pi_{q,r}(l)$ we have two 
obvious choices. The first one is
\begin{equation}\label{eq:piquenched}
\pi_{q,r}(l)\equiv\av{\frac{n_{q,r}(l)}{n(l)}}_{n(l)\neq0},
\end{equation}
where the subscript denotes that we restrict the average to the ensemble of
graphs for which $n(l)$ is not zero. The second possibility is to use
\begin{equation}
\pib_{q,r}(l)\equiv\frac{\av{n_{q,r}(l)}}{\av{n(l)}}.
\end{equation}
In Ref.~\cite{bo} we have argued that the first {\em quenched} definition
is more natural in the context of random graphs. However, it is much
more difficult to work with. In this paper we will assume that the
ensemble of generic trees is {\em self-averaging} and the two above
definitions are equivalent. For a more detailed discussion of this issue we refer
to \cite{bo}.
Similarly we define
\begin{equation}
\pib_{q}(l)\equiv\frac{\av{n_{q}(l)}}{\av{n(l)}},
\end{equation}
and the connected two point probability
\begin{equation}\label{eq:pbcon}
\pib^{con}_{q,r}(l)\equiv\pib_{q,r}(l)-\pib_q(l)\,\pib_r(l).
\end{equation}
We further define the connected correlation function \cite{bs,b1,b2}
\begin{equation}\label{eq:corrf}
\pi^{con}_{\bar q,\bar r}(l)\equiv\sum_{q,r}q \, r \,\pib^{con}_{q,r}(l).
\end{equation}
Finally we define average degree of the vertices at the distance $l$
from a vertex of degree $q$ as follows:
\begin{equation}\label{eq:def-kl}
\bar k_l(q)\equiv\frac{\av{n_{q,\bar{r}}(l)}}{\av{n_q(l)}},
\end{equation}
where
\begin{equation}
n_{q,\bar{r}}(l)\equiv\sum_{r} r \, n_{q,r}.
\end{equation}

\section{Generic random trees}
\label{sec:trees}

We consider an ensemble of all labeled trees with the probability
measure \eqref{eq:PT}.  The partition function $\Omega_V$ is defined
as the sum of the weights of all the trees in the ensemble
\begin{equation}\label{eq:weight}
\OV\equiv\frac{1}{V!}\sum_{T\in\mathcal{T}(V)}\prod_{i\in T}\w_{q_i}.
\end{equation}  
The partition function of the corresponding grand-canonical ensemble
is defined by the discrete Laplace transform
\begin{equation}\label{eq:omega-grand}
\Omu=\sum_{V=1}^\infty e^{-\mu V} \OV.
\end{equation}
We will use the
field theory approach  to calculate it \cite{jk}.  We define the function
\begin{equation}\label{eq:W}
W(\mu)\equiv\int\text{d}\phi  
\exp \left[ N \left(-\frac{1}{2}\phi^2+e^{-\mu}\sum_{q=0}\frac{\w_{q}}{q!}\phi^q\right)\right].
\end{equation}
Its formal perturbative expansion in $e^{-\mu}$ will generate Feynman's diagrams
with desired weights and symmetry factors  (for an
introduction see any textbook on field theory, e.g.,
Refs.~\cite{bdfn} and \cite{ift} or Ref.~\cite{graphs}). This expansion will,
however, contain all the graphs including those which are not
connected or contain loops.  We can restrict the expansion to
connected graphs only by considering the function $\log
W(\mu)$.  To obtain just the tree graphs we will use the expansion in
$N^{-1}$.  According to Feyman's rules for the expression \eqref{eq:W}
each edge in the graph introduces a factor $N^{-1}$ and each vertex a
factor $N$ which together contribute $N^{-E+V}$, where $E$ is the
number of edges in the graph. If $L$ is the number of independent
loops in the graph then $E-V = L-1$, so the first term of the $N^{-1}$
expansion will group graphs with no loops, the second one graphs with
one loop, and so on.

That means that the contribution of tree graphs is given by the first
term in the saddle-point approximation. 
The saddle-point equation is
\begin{equation}
  \frac{\text{d}}{\text{d}\phi}\left(-\frac{1}{2}\phi^2+e^{-\mu}\sum_{q=0}^\infty \frac{\w_{q}}{q!}\phi^q\right)=0.
\end{equation}
We will denote by $Z(\mu)$ the
solution of the above  equation and rewrite it as
\begin{equation}\label{eq:rec}
Z(\mu)= e^{-\mu}\sum_{q=1}^\infty \frac{\w_{q}}{(q-1)!}Z^{q-1}=
e^{-\mu}\frac{F\left(Z(\mu)\right)}{Z(\mu)},
\end{equation}
where
\begin{equation}\label{eq:F}
F(Z)\equiv\sum_{q=1}^\infty \frac{\w_{q}}{(q-1)!}Z^q.
\end{equation}
\begin{figure} 
\begin{center} 
\includegraphics[width=8.5cm]{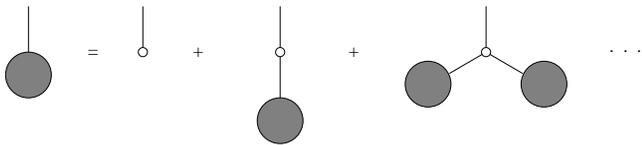} 
\end{center}
\caption{\label{fig:Z}Graphical representation of Eq.~\eqref{eq:rec}.
  Each gray bubble corresponds to the sum over planted trees given by
  the partition function $Z(\mu)$. Branches without a vertex (small empty circle) at one end
  denote a stem.}
\end{figure}
Inserting Eq.~\eqref{eq:rec} into Eq.~\eqref{eq:W} and taking the
logarithm to keep only connected graphs we obtain
\begin{equation}
\begin{split}
\label{eq:Omega}
\Omega(\mu)&=
e^{-\mu}\sum_{q=0}^\infty \frac{\w_{q}}{q!}Z^q(\mu)-\frac{1}{2}Z^2(\mu).
\end{split}
\end{equation}
It is easy to check that
\begin{equation}
e^{-\mu}Z(\mu)=\frac{\partial\Omu}{\partial \w_{1}}.
\end{equation}
Figure~\ref{fig:Z} shows a graphical interpretation of $Z(\mu)$: it
is the partition function of the ensemble of {\em planted}
trees \footnote{The trees considered in Refs.
  \cite{adfo} and \cite{bb} were planar. Here we do not impose such a
  restriction. The only difference is the appearance of the $1/(q-1)!$
  factor. This is due to the fact that now we are free to permute
  branches  emerging from the vertex.}.  Planted trees are the trees
with a {\em stem} attached to one of the vertices.  
Its properties and resulting critical
behavior were calculated in Refs.~\cite{adfo,bb,jk}. 
\begin{figure}[!b]
\begin{center}
\includegraphics[width=7.5cm]{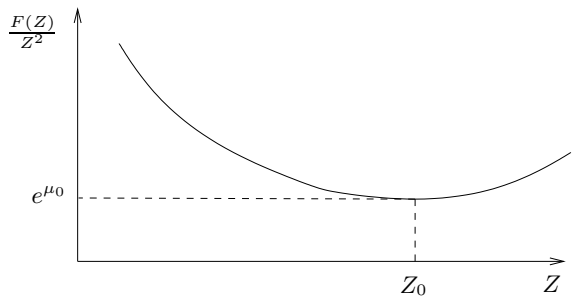}
\end{center}
\caption{\label{fig:singularity}Graphical representation of Eq.~\eqref{eq:rec-2} and the singularity of $Z(\mu)$.}
\end{figure}
The model has two main phases. In the so called {\em generic} or {\em
  tree} phase the the function $F(Z)/Z^2$ has a minimum inside its
domain (see Fig.~\ref{fig:singularity}) and so Eq.~\eqref{eq:rec}, which can be rewritten as
\begin{equation}\label{eq:rec-2}
e^{\mu}=\frac{F(Z)}{Z^2},
\end{equation}
does not have any solution for $\mu<\mu_0$. The 
function $Z(\mu)$ has a singularity at $\mu_0$ given by  the condition for the minimum
\begin{equation}\label{eq:Z0}
2F(Z_0)=Z_0F'(Z_0),\quad\text{and}\quad \mu_0=\log \frac{F(Z_0)}{Z^2_0}.
\end{equation}
At this singularity the partition  function behaves like
\begin{equation}\label{eq:treeexp}
Z(\mu)\approx Z_0-Z_1\sqrt{\mu-\mu_0}+Z_2(\mu-\mu_0),
\end{equation} 
regardless of the form of the weights $\w_{q}$. In this contribution
we will limit our self to this phase only.  Inserting the expansion
\eqref{eq:treeexp} into Eq. \eqref{eq:rec} and expanding to the order
$\Delta\mu\equiv \mu-\mu_0$ ($Z_2$ cancels in the resulting equation)
we obtain
\begin{equation}\label{eq:Z1}
\frac{Z_0^2}{Z^2_1}=\frac{1}{2}\frac{F''(Z_0)Z_0^2}{F(Z_0)}-1.
\end{equation}

\begin{figure}[!t]
\begin{center} 
\includegraphics[height=35mm]{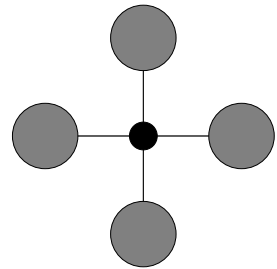}
\end{center}
\caption{\label{fig:omega_p}Graphical representation of the partition
  function $\Omega(q,\mu)$ given by Eq.~\eqref{eq:omegaq} for $q=4$.
  Each gray bubble corresponds to the sum over planted trees given by
  the partition function $Z(\mu)$; the smaller black circle represents
  the root which contributes a weight factor $e^{-\mu}w_4$; the
  additional $\frac{1}{4!}$ factor comes from the fact that the
   relative position
  of branches in the compound tree is irrelevant.  }
\end{figure}

The vertex degree distribution of this model was calculated using the
correspondence with the balls in boxes model in Ref.~\cite{bbj}.  Here
we rederive it using a different method which can be easily extended
to the case of two-point correlations studied in Ref.~\cite{b3}.  Let
us denote by $\Omega(q;\mu)$ the partition function of the rooted
grand-canonical ensemble of trees with the condition that the degree
of the root is $q$. Then
\begin{equation}\begin{split}
\label{eq:omegaq}
\Omega(q;\mu)=
\w_{q}\frac{\partial\Omu}{\partial\w(q)}=e^{-\mu}\frac{\w_{q}}{q!}Z^q(\mu).
\end{split}
\end{equation} 
The graphical interpretation of this equation is shown in
Fig.~\ref{fig:omega_p}.  The sought degree distribution is
proportional to the canonical partition function $\OV(q)$.  Inserting
the expansion \eqref{eq:treeexp} into Eq. \eqref{eq:omegaq} we obtain
\begin{eqnarray} 
\Omega(q;\mu)&\approx&e^{\mu_c}e^{\Delta\mu}\frac{\w_q}{q!}Z_0^{q}\left(1-q\frac{Z_1}{Z_0}\sqrt{\dmu}\right)\nonumber\\
&\approx&e^{\mu_c}e^{\Delta\mu}\frac{\w_q}{q!}Z_0^{q}\exp\left(
-q\frac{Z_1}{Z_0}\sqrt{\dmu}\right). 
\end{eqnarray}
The last expression has a known inverse Laplace transform
\begin{equation}\label{eq:lap}
e^{-a \sqrt{\dmu}}\; \xleftrightarrow[]{\text{Lap.}} \; 
\frac{1}{2\sqrt{\pi}}\,\frac{a}{V^{\frac{3}{2}}}\;e^{-\frac{a^2}{4V}},
\end{equation}
so finally
keeping only the first terms in the $V^{-1}$ expansion 
and fixing the normalization we obtain the formula
\begin{equation}\label{eq:piq}
\pi(q)=\frac{1}{F(Z_0)}\;\frac{\w_{q}Z_0^{q}}{(q-1)!}.
\end{equation}
Using the above formula we can give an interpretation of the
right-hand side of Eq.~\eqref{eq:Z1}
\begin{eqnarray}
\frac{1}{2}\frac{F''(Z_0)Z_0^2}{F(Z_0)}-1&=&
\frac{1}{2}\frac{\sum_{q=1} \frac{q(q-1)}{(q-1)!}w_q Z_0^q}{F(Z_0)}-1\nonumber\\
&=&\frac{1}{2}\sum_{q=1} q(q-1)\pi(q)-1\nonumber\\
&=&\frac{1}{2}\left(\av{q^2}-4\right).
\end{eqnarray}
Here we have used the fact that on trees the average degree $\av{q}=2$ (in
the large $V$ limit). Please note that
\begin{equation}
\av{q^2}-4=\av{(q-2)^2}\ge 0.
\end{equation}
The $\av{q^2}$ is equal to $4$  only in the $\omega_1\rightarrow 0$ limit.

\subsection{Correlations in generic random trees}
\label{sec:correlationstrees}

\begin{figure}[!t]
\begin{center}
\includegraphics[width=7cm]{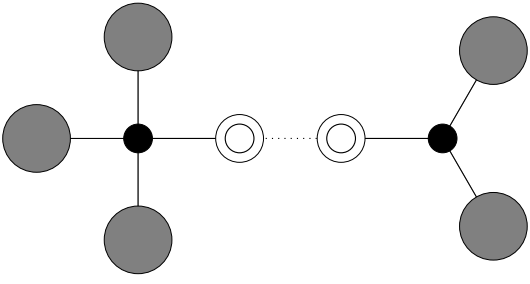}
\end{center}
\caption{\label{fig:omega_pql}Graphical representation of the
  partition function $\Omega_l(q,r;\mu)$ given by
  Eq. \eqref{eq:omega_pql} for $q=4$ and $r=3$.  Gray bubbles
  correspond to the partition function $Z(\mu)$ and the smaller black
  circles mark the vertices with degrees $q$ and $r$; double circles
  represent the $l-1$ vertices along the path connecting them in which
  we sum over all possible insertions of the $Z(\mu)$ function (see
  Fig. \ref{fig:node}).}
\end{figure}
\begin{figure}[!t]
\begin{center}
\includegraphics[width=7cm]{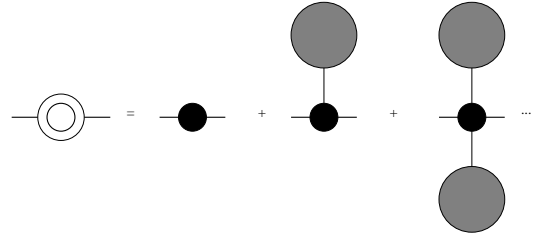}
\end{center}
\caption{\label{fig:node}Double circles denote the summation over all possible insertions of the $Z(\mu)$ function (gray bubbles).}
\end{figure}

We proceed as in the previous section but this time we introduce a
partition function $\Omega_l(q,r;\mu)$ of all the trees with two
points marked, such that the points are at the distance $l$ and have
degrees $q$ and $r$, respectively.  Because we are considering the
trees there is exactly one path linking the two marked vertices (see
Fig.~\ref{fig:omega_pql}).  As in the previous section we can express the
partition function $\Omega_l(q,r;\mu)$ by $Z(\mu)$ \cite{b2,adj}
\begin{eqnarray}\label{eq:omega_pql}
  \Omega_l(q,r;\mu)&=&\frac{e^{-\mu}\w_{q}}{(q-1)!}\;Z^{q-1}(\mu)\frac{e^{-\mu}\w_r}{(r-1)!}\;Z^{r-1}(\mu)\nonumber\\
  &\times& \left[e^{-\mu}\sum_{k=2}\frac{w_{k}}{(k-2)!}\;Z^{k-2}\right]^{l-1}.
\end{eqnarray}
The last term comes from the vertices along the path for which we have
to sum up all the possible insertions of the $Z(\mu)$ function (see
Fig.~\ref{fig:node}). This summation can be done in the following way:
\begin{eqnarray}
e^{-\mu}\sum_{q=2}\frac{w_{q}}{(q-2)!}\;Z^{q-2}&=&
e^{-\mu}\frac{\partial}{\partial Z}\sum_{q=1}\frac{w_{q}}{(q-1)!}\;Z^{q-1}\nonumber\\
&=&e^{-\mu}\frac{\partial}{\partial Z}\frac{F(Z)}{Z}.
\end{eqnarray}
Differentiating Eq.~\eqref{eq:rec} with respect to $\mu$ we come to the relation
\begin{equation}
Z'(\mu) \left(1 - e^{-\mu}\frac{\partial }{\partial Z}\frac{F(Z)}{Z}\right)= -Z.
\end{equation}
Using it we finally obtain
\begin{equation}\begin{split}
e^{-\mu}\sum_{q=2}\frac{w_{q}}{(q-2)!}\;Z^{q-2}
=
1+\frac{Z(\mu)}{Z'(\mu)}.
  \end{split}
\end{equation}
Inserting into Eq.~\eqref{eq:omega_pql} first the above formula  and then the expansion \eqref{eq:treeexp} we get
\begin{eqnarray}
  &&\Omega_l(p,q;\mu)\approx\frac{\w_q}{(q-1)!}\,\frac{\w_r}{(r-1)!}\;Z_0^{q+r-2}\nonumber\\
  &&\times\left(1-\frac{Z_1}{Z_0}\sqrt{\Delta\mu}\right)^{q+r-2}\left(1-2\frac{Z_0}{Z_1}\sqrt{\Delta\mu}\right)^{l-1}.
\end{eqnarray}
This can be further approximated by
\begin{eqnarray}
\Omega_l(p,q;\mu)&\approx& \frac{\w_q}{(q-1)!}\,\frac{\w_r}{(r-1)!}\;Z_0^{q+r-2}\nonumber\\
&\times& e^{-\left[\frac{Z_1}{Z_0}(q+r-2) + 2\frac{Z_0}{Z_1}(l-1)\right]\sqrt{\Delta\mu}},
\end{eqnarray}
and using Eq. \eqref{eq:lap} we obtain to the leading order in $V$
\begin{eqnarray}
\av{n_{qr}}\propto\Omega_l(p,q;V)&\approx&\frac{\w_q}{(q-1)!}\,\frac{\w_r}{(r-1)!}\;Z_0^{q+r-2}\nonumber\\
&\times& \left[\frac{Z_1}{Z_0}(q+r-2) + 2\frac{Z_0}{Z_1}(l-1)\right].\nonumber\\
\end{eqnarray}
Finally, we get
\begin{align} 
\pib_{q,r}(l)=&\,\pi_q \pi_r \frac{ (q+r-2) +\left(\av{q^2}-4\right)(l-1)} {2 + \left(\av{q^2}-4\right)(l-1)}, \label{eq:piqrl}\\
\pib_{q}(l)= &\,\pi_q \frac{ q + \left(\av{q^2}-4\right)(l-1)} {2 + \left(\av{q^2}-4\right)(l-1)}.\label{eq:piql} 
\end{align}
Inserting this into Eqs. \eqref{eq:pbcon} and \eqref{eq:def-kl} we
obtain the results \eqref{eq:pbcontree} and \eqref{eq:kl}. Summing up   
Eq.~\eqref{eq:pbcontree} over $q$ and $r$ we get the 
connected correlation function \eqref{eq:corrf}
\begin{equation}
\begin{split}
\pib^{con}_{\bar q,\bar r}(l)=-\frac{\left[\av{q^2}-4\right]^2}{\left[2+ (\av{q^2}-4)(l-1)\right]^2}.
\end{split}
\end{equation}

\subsection{Example 1}
\label{sec:erdos}

In the first example we put $\w_{q}=1$, so all the trees in the
ensemble have the same weight. In this case $F(Z)=Z e^{Z}$.  The
solution of Eq.  \eqref{eq:Z0} is $Z_0=1$  from which
follows:
\begin{equation} 
\pi_q=\frac{1}{e}\frac{1}{(q-1)!},\quad
\av{q^2}=\sum_{q=1}^\infty\frac{1}{e}\frac{q^2}{(q-1)!}=5.
\end{equation}
leading to
\begin{align}
\bar{\pi}_q(l)&=\pi_q\frac{q+l-1}{1+l},\\
\pib^{con}_{q,r}(l)&=-\frac{1}{e^2}\frac{(q-2)(r-2)}{(l+1)^2}
\frac{1}{(q-1)!}\frac{1}{(r-1)!},
\end{align}
and
\begin{equation}
\bar \pi^{con}_{\bar q, \bar r}(l)=-\frac{1}{(l+1)^2},\quad \bar k_{l}(q)=2+\frac{1}{q+l-1}.
\end{equation}

\subsection{Example 2: Scale-free trees}
\label{sec:sf}

In this example we choose $\w_{q}=q^{-\beta}(q-1)!$ which corresponds
to the planar graphs studied in Refs.~\cite{adfo} and \cite{bb}. Then $F(Z)$ is
given by the polylogarithm function $\Li_\beta(Z)$
\begin{equation}
F(Z) = \sum_{q=1}^\infty \frac{Z^q}{q^\beta} \equiv \Li_\beta(Z).
\end{equation}
and Eq.~\eqref{eq:Z0} takes the form
$2\Li_\beta(Z_0)=\Li_{\beta-1}(Z_0)$.  It has the solution for
$\beta<\beta_C$ with $\beta_C\approx2.4788$ given by
$2\zeta(\beta_C)=\zeta(\beta_C-1)$.  At the critical value of
$\beta=\beta_C$ the partition function no longer scales as in Eq.~\eqref{eq:treeexp}
and in principle we cannot use the Laplace transform Eq.~\eqref{eq:lap} anymore.
However, as shown in Ref.~\cite{bepw} the large $V$
behavior is not changed and we expect our formula to hold in the
large $V$ limit. From Eq.~\eqref{eq:piq} we read-off the degree
distribution
\begin{equation}\label{eq:piscale}
\pi(q)=\frac{q^{-\beta}Z_0^{q-1}}{F(Z_0)}.
\end{equation} 
 At the critical value of $\beta$,
$Z_0=1$ and the vertex degree distribution is scale free. The average
\begin{equation}
  \av{q^2}=2+\frac{1}{\Li_\beta(Z_0)}\left[\Li_{\beta-2}(Z_0)-\Li_{\beta-1}(Z_0)\right],
\end{equation}
diverges as $\beta\rightarrow\beta_C$. Formula
\eqref{eq:pbcontree} leads for $l>1$ to
\begin{equation} \label{eq:piconsf}
\lim_{V\rightarrow\infty}\bar\pi^{con}_{q,r}(l)=0.
\end{equation}
This would imply that the correlations vanish in the large $V$ limit. However, this
limit \eqref{eq:piconsf} is not uniform. It is easy to check that the {\em integrated} correlation functions do  not disappear
\begin{equation} \label{eq:pisf}
\lim_{V\rightarrow\infty}
\bar \pi^{con}_{\bar q,\bar r}(l)=-\frac{1}{(l-1)^2},
\end{equation}
and
\begin{equation}\label{eq:klsf}
\lim_{V\rightarrow\infty}\bar k_l(q)=2+\frac{1}{l-1}.
\end{equation}
Please note that the above results are universal and valid for any kind of
scale-free trees  with $\beta<3$. 

\subsection{Monte Carlo simulations}
\label{sec:simulations}

The results obtained in the previous sections are valid only in the strict
$V\rightarrow\infty$ limit and it is clear that for  finite $V$  
 our formulas will not hold for any $l$. Defining the
average distance on a graph
\begin{equation}
\av{l}=\sum_l l\frac{n(l)}{V^2},
\end{equation}
we may expect that the formulas will be valid only for $l\ll \av{l}$.
The scaling of $\av{l}$ with the graph size depends on the Hausdorff's
dimension $d_H$
\begin{equation}
\av{l}\sim V^{{1}/{d_H}}.
\end{equation}
For generic trees considered here $d_H=1/2$. For scale-free
trees considered in example 2 we expect
\begin{equation}
d_H=\frac{1}{\gamma}, \qquad
\gamma=\frac{\beta_C-2}{\beta_C-1},
\end{equation}
which gives $d_H\approx 3$. In the case of scale-free trees the volume dependence 
manifests itself by the cutoff in the degree distribution $\pi_q$ as well \cite{bck}.

Expecting finite size effects to be more severe in the
scale-free trees, we checked the $V$ dependence performing
MC simulations of the ensemble described in the
example~2. We have used an algorithm similar to ``baby-universe
surgery'' \cite{abbjp}. The basic move consisted of picking up an edge
at random and cutting it. Then the
smaller of the two resulting trees was grafted on some vertex of the bigger one. The most
time consuming part of the algorithm was to find which tree was
smaller. To save time the two trees were traversed simultaneously until one of them was filled
completely.  Additionally, to pick the attachment point from the bigger tree efficiently, the vertices of the trees were marked during the traversal.
This move was supplemented with moves consisting of
cutting up leaf nodes and attaching them to some other parts of the
tree. This was much faster as it did not require traversing the tree.
However, the autocorrelation time for such moves alone was
much higher, especially for the scale-free trees. Because those trees
are at the phase transition between the generic and the crumpled
phase \cite{bb} the autocorrelation time is high even for the tree
grafting algorithm. 

\begin{figure}[!t]
\includegraphics[height=7cm]{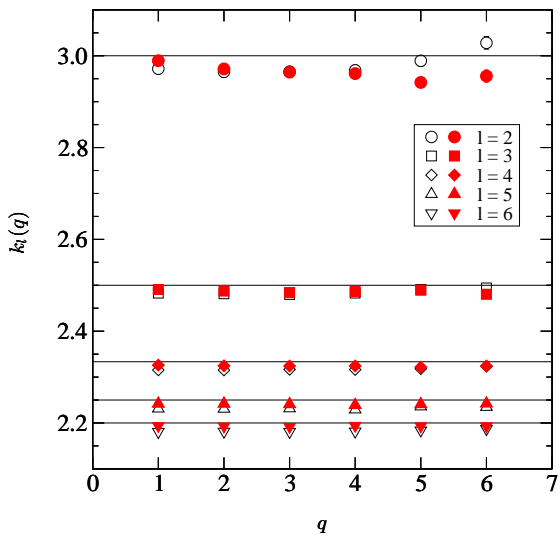}
\caption{\label{fig:kl} Average degree of distance $l$ neighbors $k_l(q)$ for $l=2$ to $6$ and trees with
  64~000 (empty symbols) and 128~000 (filled symbols) vertices. Each symbol denotes
  different $l$; straight lines are the predictions given by Eq.~\eqref{eq:klsf}.}
\end{figure}

\begin{figure}[!t]
\includegraphics[height=7cm]{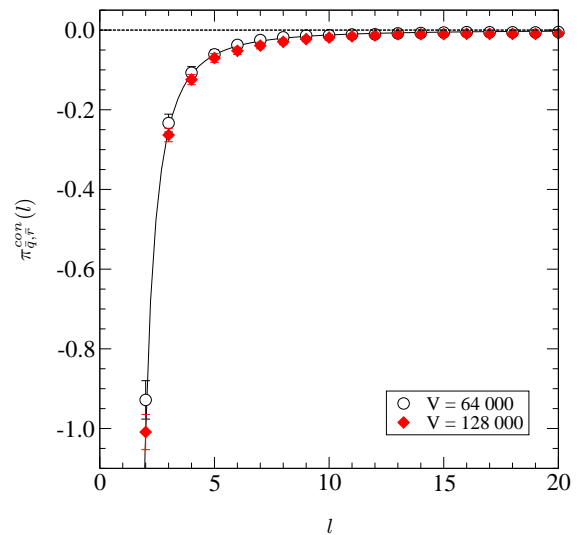}
\caption{\label{fig:conn} Connected correlation function $\pi_{\bar q,\bar
    r}^{con}(l)$ for scale-free trees with 64~000 (empty circles) and
  128~000 (filled diamonds) vertices. The solid line represents the prediction given by Eq.~\eqref{eq:pisf}.}
\end{figure}

We have simulated trees of the size up to 128~000 vertices. To verify
to which extent the ensemble is self-averaging we measured the
quenched quantities which we then compared to our predictions.  In
Fig.~\ref{fig:kl} we have plotted the measured
\begin{equation}
k_l(q)=\av{\frac{n_{q,\bar r}(l)}{n_q(l)}},
\end{equation} 
as a function of $q$ for various values of $l$.  Please note the large
finite size effects for $l=2$. This is to be expected: For finite $V$ the
$\av{q^2}$ is also finite and actually grows slowly with
$V$ \cite{bck}. For larger $l$ the agreement with our results  is quite good. 
Figure~\ref{fig:conn} shows the quenched correlation function
\begin{equation}
\pi^{con}_{\bar q,\bar r}(l)=
\av{\frac{n_{\bar q,\bar r}(l)}{n(l)}}-
\av{\frac{n_{\bar q}(l)}{n(l)}}^2,
\end{equation}
which is also well reproduced by our results. 

\section{Summary and Discussion}
\label{sec:summary}

The appearance of the long-range correlations in generic trees is
puzzling. Usually we expect the powerlike (scale-free) behavior to
be manifested in systems at the criticality. The trees studied here
apart from the scale-free example are, however, not critical. The
free-energy density can be calculated in the infinite volume limit
and remains an analytic function of the weights $\omega_q$
\cite{bb,bbj}.  It has been also shown that the critical behavior in
random trees is not associated with the diverging correlation length
\cite{bbju}. The correlations described here are thus of the
structural and not of dynamical origin. A possible mechanism
explaining it was proposed in Refs. \cite{b1} and \cite{bo}: in connected graphs
vertices of degree one must have neighbors of degree greater than
one. It remains, however, to be understood how this effect can be
propagated to larger distances.

\acknowledgments

The authors thank Z.~Burda for valuable discussion. This paper was
partially supported by EU grants No. MTKD-CT-2004-517186 (COCOS) and
No. MRNT-CT-2004-005616 (ENRAGE). P.B. thanks the Service de Physique
Th\'eorique, CEA/Saclay for the kind hospitality during his stay. 

\appendix*

\section{Maximal  entropy}

For each choice of the weights $\w_q$ and a given number of vertices
$V$ the ensemble \eqref{eq:PT} has a well defined degree distribution
$\pi(q)$.  Asymptotically for large $V$ this distribution is given
by Eq.~\eqref{eq:piq}. Ref. \cite{BauerBernard} contains a
proof that the probability measure \eqref{eq:PT} has maximal entropy
among all the measures producing the distribution
$\pi(q)$. Here we repeat their arguments for completeness. 

We start with an expression for the entropy plus the necessary
Lagrange multipliers to force the constraints
\begin{eqnarray}\label{eq:max-entropy}
\cal{S}&=&-\frac{1}{V!}\sum_{T\in\mathcal{T}} P(T) \log P(T) + \frac{\lambda}{V!} \sum_{T\in\mathcal{T}} P(T)\nonumber\\
&&+\sum_{q=1}^V\frac{\lambda_q}{V!}\left( \sum_{T\in\mathcal{T}} n_q(T) P(T)-\pi(q) V \right).
\end{eqnarray}
Differentiating the above with  respect to $P(T)$ 
we get
\begin{equation} 
\log P(T)=\lambda-1+\sum_{q}\lambda_q n_q(T),
\end{equation}
leading to
\begin{equation} 
P(T)=e^{\lambda-1}\prod_{q}e^{\lambda_q n_q(T)}.
\end{equation}
Putting 
\begin{equation}
e^{\lambda-1}=\frac{\Omega_V}{V!},\quad\text{and}\quad
e^{\lambda_q}=\w_q
\end{equation}
we obtain Eq.~\eqref{eq:PT}. We will now prove that this
measure is a unique solution of Eq. \eqref{eq:max-entropy}
satisfying the constraints, at least in the
$V\rightarrow\infty$ limit.  Let us assume that we have another set of
weights $\tilde{\w}_q$ that produces the same probability distribution
$\pi(q)$ \eqref{eq:piq}
\begin{equation}
\pi(q)=\frac{1}{F(Z_0)}\frac{\w_q Z_0^q}{(q-1)!}
=\frac{1}{F(\tilde{Z}_0)}\frac{\tilde{\w}_q \tilde{Z_0}^q}{(q-1)!}.
\end{equation}
It follows that: 
\begin{equation}
\tilde{\w}_q=
\frac{F(\tilde{Z}_0)}{F(Z_0)}\left(\frac{Z_0}{\tilde{Z}_0}\right)^q\;w_q,
\end{equation}
hence
\begin{eqnarray}
\prod_{i\in T}\tilde{\w}_{q_i}&=&
\prod_{i\in T}\frac{F(\tilde{Z}_0)}{F(Z_0)}\left(\frac{Z_0}{\tilde{Z}_0}\right)^{q_i}w_{q_i}\nonumber\\
&=&\left(\frac{F(\tilde{Z}_0)}{F(Z_0)}\right)^V\left(\frac{Z_0}{\tilde{Z}_0}\right)^{\sum_{i\in T}q_i}\prod_{i\in T}w_{q_i}.
\end{eqnarray}
But this gives identical probability measure to Eq.~\eqref{eq:PT} because of the condition $\sum_{i\in T}q_i=2V-2$ valid for each tree $T$.

\end{document}